\documentclass[aps,amssymb,amsmath,pre,twocolumn,superscriptaddress]{revtex4-1}

\usepackage{graphicx}
\usepackage{epstopdf}
\usepackage{amssymb}
\usepackage{epstopdf}
\usepackage{amsmath}
\usepackage{amsthm}
\usepackage[extension=xxx]{hyperref}
\usepackage{color}
\usepackage{listings}

\usepackage[toc,page]{appendix}

\def\be{\begin{equation}}   \def\ee{\end{equation}}
\def\eq#1{{Eq.\ref{#1}}}    \def\fig#1{{Fig.\ref{#1}}}

\begin{document}

\title{Cooperative protein transport in cellular organelles}

\author{S. Dmitrieff}
\email{serge.dmitrieff@espci.fr}
\author{P. Sens}
\email{pierre.sens@espci.fr}
\homepage{http://www.pct.espci.fr/~pierre}
\affiliation{Laboratoire Gulliver (CNRS UMR 7083)\\ 
ESPCI, 10 rue Vauquelin, 75231 Paris Cedex 05\\
 France}

\date{\today}

\begin{abstract}
Compartmentalization into biochemically distinct organelles constantly exchanging material is one of the hallmarks of eukaryotic cells. In the most naive picture of inter-organelle transport driven by concentration gradients, concentration differences between organelles should relax. We determine the conditions under which cooperative transport, i.e. based on molecular recognition, allows for the existence and maintenance of distinct organelle identities. Cooperative transport is also shown to control the flux of material transiting through a compartmentalized system, dramatically increasing the transit time under high incoming flux. By including chemical processing of the transported species, we show that this property provides a strong functional advantage to a system responsible for protein maturation and sorting.\\

\end{abstract}

\maketitle

Eukaryotic cells contain many specialized compartments (organelles) constantly exchanging molecules through complex energy-consuming transport processes.  Along the secretory pathway for instance, proteins and lipids synthesized in the endoplasmic reticulum (ER) are transported to the Golgi complex (itself sub-compartmentalized into stacked cisternae) for maturation and sorting, before being conveyed to particular cellular locations \cite{lippincott:2000}. ER and Golgi (and even the different Golgi cisternae) are characterized by fairly different lipid and protein compositions \cite{wilson:2010}. There is a widespread interest in understanding the molecular and physical bases permitting these organelles to maintain their {\em identity}, namely their specific chemical composition, while  constantly exchanging material\cite{misteli:2001}. 

In a system where fluxes are linearly related to concentration differences ({\em i.e.} satisfying Fickian diffusion), stationary concentration  gradients can only be maintained by external fluxes.
Fluctuations of the fluxes yield fluctuations of the local composition, and robust compartment identity (namely the existence of stable concentration heterogeneity) is not to be expected. In the secretory pathway however, transport is heavily influenced by molecular recognition: transported molecules, carriers and recipient organelles cooperate through complex networks of molecular interactions \cite{chen2001snare,caviston2006microtubule}. The goal of this paper is to study the consequences of such cooperative transport on the stationary state of a simple (2-compartment) system, and on its possible function in protein maturation and sorting. Our model generalizes and extends the vesicular transport model proposed by Heinrich and Rapoport \cite{heinrich:2005}, where the rates of vesicle exchange between compartments are influenced by their composition. We discuss below both the case of a closed system and of an open system with incoming and outgoing fluxes. By including chemical transformation of the transported species, we show that cooperative transport can strongly increase the accuracy of a system responsible for protein maturation and sorting.

\section{Stationary compartment differentiation in a closed system}\label{sec-closed}

 We first consider a single protein species distributed between two compartments constantly exchanging material, and indicate how these results might be extended to a multicomponent system. 
 
 \subsection{One species system}\label{sec-closed_1species}
 
 \begin{figure*}[t]
\includegraphics[width=15cm]{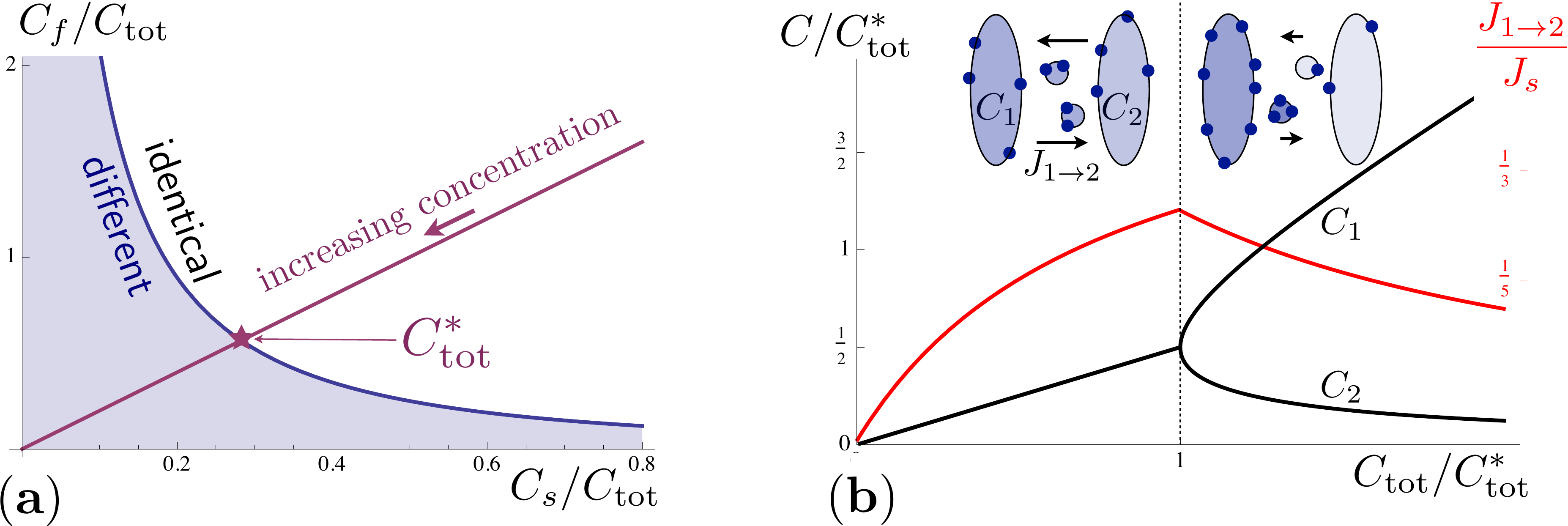}
\caption{\label{closed} {\bf (a)} Location of the critical region in the parameter  space  $\{C_s/C_{\rm tot},C_f/C_{\rm tot}\}$ where stationary compartment differentiation occurs in a closed system (shaded blue, \eq{cstar}). Increasing the total concentration  $C_{\rm tot}$ moves the system along the red line (arrow).  {\bf (b)}: Variation of the stationary compositions (black) and flux (red) with the total concentration, showing the breaking of symmetry for $C_{\rm tot}>C^*_{\rm tot}$ (\eq{cstar}).}
\end{figure*}
 
 We assume below that the total mass of the system (and the mass of each compartment) is maintained constant by  an unspecified regulatory mechanism, so that the evolution of the concentration $C_1$ in compartment $1$  can generically be described by the Master equation \cite{vankampen:2007}:
\begin{eqnarray}
\partial_t C_1  =I_1-J_{1 \rightarrow 2}(C_1,C_2)+J_{2 \rightarrow 1}(C_2,C_1)
\label{Cdot}
\end{eqnarray}
with a similar expression for compartment $2$ obtained by the transformation $1\leftrightarrow2$. Here, $J_{\alpha \rightarrow \beta}(C_\alpha,C_\beta)$ is the mean flux  from compartment $\alpha$ (with concentration $C_\alpha$) to compartment $\beta$ (of concentrations $C_\beta$). Compartments will naturally reach different concentrations if they follow different exchange rules. We focus on the more interesting case where  $J_{1\rightarrow 2}(C_1,C_2)=J_{2\rightarrow 1}(C_1,C_2)$.  The source and sink term $I_1$  in \eq{Cdot} may  include both external fluxes in and out of compartment $1$ and chemical transformation within this compartment. 

The transport of cargo between organelles may be separated into three distinct steps; {\em step $1$}: cargo packaging inside a membrane-based  carrier, such as a small protein-coated vesicle or a membrane  tubule \cite{bremser1999coupling}, {\em step $2$}: the actual transport between secreting and receiving compartments, often involving molecular motors moving along cytoskeletal filaments\cite{cole1995organization,welte2004bidirectional}, and {\em step $3$}: fusion of the carrier with the receiving organelle (see \cite{bonifacino:2004} for a review).
For {\em step $1$}, we call $J_s$ the total flux of material secreted by a compartment, and $S$ the fraction of this flux (a number between $0$ and $1$) occupied by the species of interest. Let us assume {\em step $2$} to be infinitely fast (we relax this hypothesis in Sec.\ref{sec-closed_finitetime}). In this case,  any vesicle secreted by compartment $1$ immediately merges ({\em step $3$}) either with compartment $2$, with a probability  $P_{1\rightarrow2}$, or back with compartment $1$ (with probability  $P_{1\rightarrow1}=1-P_{1\rightarrow2}$). The mean flux from compartment $1$ to compartment $2$ may thus be written:
\begin{equation} 
J_{1 \rightarrow 2}(C_1,C_2) = J_s(C_1) S(C_1) P_{1 \rightarrow 2}(C_1,C_2).
\label{jab}
\end{equation}

For a closed system with fixed concentration $C_{\rm tot}=C_1+C_2$ (no source and sink term), the symmetric state: $C_1=C_2=C_{\rm tot}/2$ is a stationary solution ($\partial_t C_1=\partial_t C_2=0$). Linear stability analysis \cite{vankampen:2007} shows that the symmetric solution is {\em unstable} provided $\left( \partial_{C_1} J_{1 \rightarrow 2} \right)_{C_1=C_2=C_{\rm tot}/2} < 0$.  The case of particles randomly entering transport vesicles which are secreted at constant rate and fuse non-specifically with either compartment corresponds to a linear flux ($J_{1\rightarrow2} \sim C_1$) (akin to passive diffusion) and leads to identical compartments.
Spontaneous compartment differentiation can only occur if the flux reaching the second compartment {\em decreases} with increasing concentration in the first one, and this requires non-linear transport ({\em cooperativity}).

Let us assume the fluxes to have two rather universal types of non-linearity as a function of concentration. Firstly, the out-going flux of a given species should saturate at high species concentrations. This can be due to  the limited capacity of transport vesicles, the limited availability of vesicle-coating proteins, or the formation of aggregates inapt for transport in a compartment beyond a critical concentration. For simplicity, we choose to keep the flux of secreted vesicles constant  (and write it $J_s\equiv K_0 C_s$), although direct interactions between cargoes and coat proteins are known to exist \cite{bremser1999coupling}. The packaging  fraction $S$ is assumed to saturate beyond a  concentration $C_s$ following a Michaelis-Menten saturation \cite{michaelis1913kinetik}:
\begin{equation}
S (C_{1}) =\frac{C_{1}}{C_{1}+C_s}
\label{saturation}
\end{equation}

Secondly, vesicle fusion is known to be strongly regulated by specific molecular interactions,  including, but not restricted to, interactions between matching pairs of SNAREs\cite{chen2001snare}. Quantitative models have  shown the importance of this step for the generation and maintenance of non-identical compartments, using fairly detailed mathematical modeling of the different pairs of SNAREs \cite{heinrich:2005,binder:2009} and/or extensive numerical simulations \cite{kuhnle:2010}. Numerous factors can however influence the delivery of transport vesicles, including specific interactions between the cargo and molecular motors \cite{caviston2006microtubule}. Here, we adopt a very generic treatment of specific fusion, where the fusion probability $P_{1\rightarrow 2}$ deviates from its nonspecific value because of two-body interactions between constituents of the vesicle and the receiving compartment: $P_{1\rightarrow2} - 1/2\sim S(C_1) C_2$. After normalization, the  probability may be written:
\begin{equation}
P_{1\rightarrow2}=\frac{ C_f+S (C_{1}) C_{2} }{2 C_f+S(C_{1})(C_{1} + C_{2}) }
\label{rates}
\end{equation}
where $C_f$ is the typical concentration beyond which specific fusion  becomes relevant.
Within the description outlined in Eqs.(\ref{saturation},\ref{rates}), linear transport corresponds to both characteristic concentrations being very large: $C_s,C_f\gg C_{\rm tot}$.

Spontaneous symmetry breaking (enrichment of one compartment at the expense of the other) occurs when  $\left( \partial_{C_1} J_{1 \rightarrow 2} \right)_{C_{\rm tot}/2}<0$. As shown in \fig{closed}a, this always happens at high enough concentration $C_{\rm tot}>C_{\rm tot}^*$, with 
\be
C_{\rm tot}^*{}^3=4C_sC_f(C_s+C_{\rm tot}^*)
\label{cstar}
\ee
Beyond this threshold, any small perturbation from the symmetric state brings the compartments into a stable asymmetric steady-state. As a consequence, the concentration of the least concentrated compartment (compartment $2$, say) and the flux $J_{1\rightarrow2}$ of material exchanged between compartments both decrease with increasing concentration when $C_{\rm tot}>C_{\rm tot}^*$, \fig{closed}b. At high concentration, the asymptotic solution reads $C_2\sim 2C_f C_s/C_{\rm tot}$.

Although the actual location of the critical line defined by \eq{cstar} depends on the model  (Eqs.(\ref{saturation},\ref{rates})) for the exchange flux $J_{1\rightarrow 2}$ (\eq{jab}), its existence does not. This critical behaviour is very general and stems from the presence of two competing effects : cooperative fusion promotes protein enrichment (and increases with decreasing $C_f$), while saturation of protein packaging (beyond a composition $C_s$) limits transport. Including the presence of different types of coat and fusion proteins does not fundamentally alter this picture \cite{heinrich:2005}. 

\subsection{Extension to a ${\bf n}$-species system}\label{sec-closed_nspecies}

Extending the analysis presented above to a $n$-component system is rather straightforward. Let us call $C_\alpha^i$ the concentration of the species $i$ in the compartment $\alpha$ ($\alpha=1,2$). The concentration of all species in compartment $\alpha$ can be defined as a vector $\boldsymbol{C}_{\alpha} = [C_\alpha^1,C_\alpha^2,...,C_\alpha^n]$, and satisfies the Master equation:
\begin{eqnarray}
\partial_t C_\alpha^i  =I^i_\alpha-J^i_{\alpha \rightarrow \beta}+J^i_{\beta \rightarrow \alpha}
\label{Cdoti}
\end{eqnarray}
where $J_{\alpha \rightarrow \beta}^i$ is the mean flux of the species $i$ from the compartment $\alpha$ to the compartment $\beta$, and $I_\alpha^i$ is a net source and sink term including both the presence of external fluxes of species $i$ in and out of compartment $\alpha$, and chemical transformation involving species $i$ in compartment $\alpha$.

For a closed system (no source and sink term), the total concentration for the $i$-th specie is fixed:  $C^i_{\rm tot} = C^i_{\alpha} + C^i_{\beta}$. All the equations may thus be written for the fractions  $\phi^i_{\alpha}= C^i_\alpha/C^i_{\rm tot}$, satisfying $\phi^i_{1} + \phi^i_{2} = 1$. Then $\boldsymbol{ \phi_2} = \boldsymbol{1} -\boldsymbol{\phi_1}$ becomes implicit and the master equation is now written only as a function of $\boldsymbol{\phi} \equiv \boldsymbol{\phi_1}$ :
\begin{eqnarray}
\partial_t{\phi^i}  = -j^i_{1 \rightarrow 2}(\boldsymbol{\phi}, \boldsymbol{1}-\boldsymbol{\phi})+j^i_{2 \rightarrow 1}(\boldsymbol{1}-\boldsymbol{\phi},\boldsymbol{\phi})
\label{phidot}
\end{eqnarray}
with the normalized fluxes  $j^i_{\alpha\rightarrow\beta}=J^i_{\alpha\rightarrow\beta}/C^i_{\rm tot}$. Assuming as before that both compartments follow identical exchange rules,  $\boldsymbol{\phi}_{1/2}=1-\boldsymbol{\phi}_{1/2}=[\frac{1}{2},\frac{1}{2},...,\frac{1}{2}]$ is a stationary solution. The linear stability of the symmetric solution is determined by the  Jacobian matrix ${\bf M}$ :
\begin{equation}
M_{i,k} =  - 2 \left( \partial_{\phi^{i}} j^k_{1 \rightarrow 2} \right)_{\boldsymbol{\phi}_{1/2}}
\label{M}
\end{equation}
The symmetric state is {\em unstable}, and spontaneously evolves towards a non-symmetric state if If ${\bf M}$ has at least one positive eigenvalue.

In a multi-component system, the fluxes can be written similarly to the main text: 
\begin{equation} 
J_{\alpha \rightarrow \beta}^i = J_{\alpha} (\boldsymbol{C}_{\alpha}) S^i_\alpha (\boldsymbol{C}_{\alpha}) P_{\alpha \rightarrow \beta}(\boldsymbol{C}_{\alpha},\boldsymbol{C}_{\beta})
\label{jabi}
\end{equation}
The functions $J_{\alpha}$, $S^i_\alpha$ and $P_{\alpha \rightarrow \beta}$ may contain various non-linearities. In particular $P_{\alpha \rightarrow \beta}$ may involve any combination of pair interactions $\{S_\alpha^i,C_\beta^j\}$ which can lead to a very rich behaviour. One could in particular describe in this way the transport of proteins directly interacting with the secretion (coat proteins) or the fusion (SNAREs) machinery.

\subsection{Influence of a finite vesicle fusion time}\label{sec-closed_finitetime}

If vesicular transport between secreting and receiving compartment (the so-called {\em step} 2 in \ref{sec-closed_1species}) is not infinitely fast, vesicles will dwell for some time  the inter-compartment region, and will have a complex distribution of concentration, reflecting the concentration of the emitting compartment at the time of their secretion. While this situation appears much more complex than the one described above, we now show, restricting ourselves to a one-species system for simplicity, that a model with inter-compartment dwelling of vesicles can be mapped to the simpler model with immediate fusion of vesicles.

Each vesicle can carry a given amount of proteins, and a vesicle budding from or merging with a compartment will change the concentration of this compartment. Let us call $C_v$ the resulting change of concentration in the compartment. Allowing vesicles to dwell between compartments for a finite time causes the total number of molecules in the compartments to decrease, and hence yields an effective total concentration $C_{\rm eff}$ lower than the actual total concentration in the system $C_{\rm tot}$:
\begin{equation}
C_{\rm eff} = C_{\rm tot} - \sum_{i=1}^{N_v} C_v^i
\end{equation}
where $N_v$ is the number of vesicles between compartments, and $C_v^i$ the concentration carried by the $i$-th vesicle.

This sum over all the vesicles is actually a random variable, but its mean can be computed analytically in certain cases. To know where the symmetric/asymmetric transition occurs, we may compute $C_{\rm eff}$ in a symmetric system, and the result will be valid up until the transition. If the system is symmetric, vesicles have the same average concentration $\bar{C_v}$ and $\sum_{i=1}^{N_v} C_v^i \approx N_v \bar{C}_v$. Let us assume each vesicle in the inter-cisternal medium has a rate of fusion $W_r$ towards any of the compartments. The average number of vesicles in the media is then $2 K_v / W_r$ (where $K_v$ is the rate of individual vesicle secretion). Moreover, the average vesicle concentration $\bar C_v$ can be written as the maximum concentration $C_v^{max}$ a vesicle may carry, times the average vesicle saturation fraction $\bar S$ (obtained from \eq{saturation}), leading to $C_{\rm eff} = C_{\rm tot} - 2 \frac{K_v}{W_r}C_v^{max}\bar S$.  
Finally, the product $K_v C_v^{max}$ is the number of vesicle leaving a compartment per unit time multiplied by the maximum concentration of each vesicle, and can be identified with $J_s\equiv K_0 C_s$. The critical point of  a system of total concentration $C_{\rm tot}$ with vesicles staying a finite time between the compartments can thus be obtained from the critical point (\eq{cstar})  of a system with infinitely fast fusion, but with an effective total concentration $C_{\rm eff}^{\rm sym}$ given by: 
\begin{equation}
C_{\rm eff}^{\rm sym}= C_{\rm tot} - 2 C_s \frac{K_0}{W_r} \bar S\quad,\quad  \bar{S} = \frac{C_{\rm eff}^{\rm sym}}{C_{\rm eff}^{\rm sym}+2 C_s}
\label{ceffsym}
\end{equation}
namely
\begin{eqnarray}
\frac{C_{\rm eff}^{\rm sym}}{C_{\rm tot}} = \frac{1}{2} - \frac{1+w_r}{w_r}\phi_s + \sqrt{ 2 \phi_s + \left( \frac{1}{2}  - \phi_s \frac{1+w_r}{w_r} \right)^2}\ \ \ \ \ 
\label{ceff}
\end{eqnarray}
with $\phi_s = C_s / C_{\rm tot}$, $w_r = W_r / K_0$.

 Similarly, one can compute the effective concentration in a fully asymmetric system, in which one compartment has a concentration $C_{\rm eff}^{\rm asym}$ and the other has a concentration close to zero:
\be
C_{\rm eff}^{\rm asym}= C_{\rm tot} - C_s \frac{K_0}{W_r} S' \quad,\quad  S'= \frac{C_{\rm eff}^{\rm asym}}{C_{\rm eff}^{\rm asym} + C_s}
\label{finitevestime1}
\ee
The difference between \eq{ceff} and \eq{finitevestime1} being that in the latter case, the empty compartment is sending out empty vesicles, and only vesicles from the first compartment contribute to the depletion effect. We find :
\begin{equation}
C_{\rm eff}^{\rm asym} (\phi_s,w_r) = C_{\rm eff}^{\rm sym} (\frac{1}{2} \phi_s,  w_r )
\label{finitevestime2}
\end{equation}

\begin{figure}[t]
\includegraphics[width=8cm]{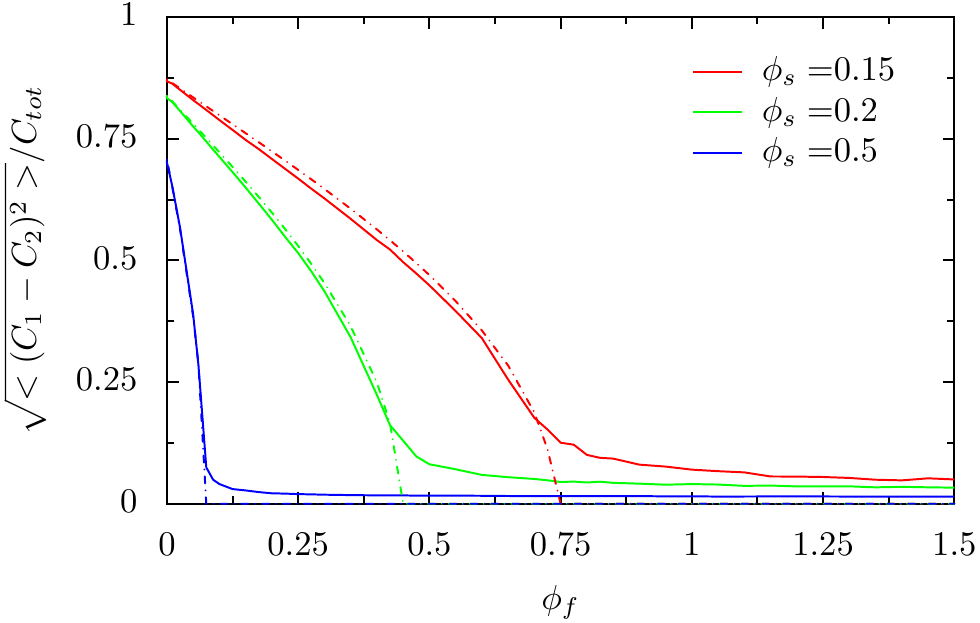}
\caption{\label{finite_ves_time} Root mean square (RMS) of the difference of concentration between the two compartments as a function of $\phi_f=C_f / C_{\rm tot}$ for various values of $\phi_s=C_s/C_{\rm tot}$. Dash-dotted lines represent the mean-field values of the concentration (normalized by $C_{\rm eff}^{\rm asym}/C_{\rm tot}$ with effective parameters $\phi_s^{eff}=C_s/C_{\rm eff}^{\rm sym}$ and $\phi_f=C_f/C_{\rm eff}^{\rm sym}$. Solid lines represent the simulated results with vesicles in the inter-compartments medium, with a vesicle fusion rate $W_r = K_0$, i.e. up to $40 \%$ of the molecules are out of the compartments. The non-zero value of the RMS in the symmetric state is due to fluctuations.}
\end{figure}

These predictions can be tested by numerical simulations (described in the Appendix). Comparison with the solution of the infinitely fast fusion model  are shown in \fig{finite_ves_time}. Not only the location of the critical line, but also the actual values of the concentrations in each compartment in the asymmetric steady state, where found to agree very well, even for low vesicle fusion rate. 

\section{Compartment differentiation in an open system}\label{sec-open}

The relative simplicity of the model presented in Sec.\ref{sec-closed_1species}, essentially characterized by two parameters ($C_s/C_{\rm tot}$ and $C_f/C_{\rm tot}$, \fig{closed}.a), allows us to address issues of direct biological relevance, such as the presence of external fluxes of material, and the possibility for chemical transformation within the system. Organelles such as the Golgi are strongly polarized, with distinct  entry and exit faces.
We investigate the consequences of cooperative transport in such open systems assuming that a particular species enters the system through compartment $1$, and exits through compartment $2$, while exchange between the two compartments proceeds as described previously. Mathematically, this amounts to including a source term $I_1=J_{in}$ and a sink term $I_2=-J_{out}$ in \eq{Cdot} yielding:
\be
J\equiv J_{1\rightarrow2}-J_{2\rightarrow 1}=J_{in}-\partial_tC_1 =\partial_tC_2+K_{\rm off}C_2
\label{kineqc}
\ee
where a simple linear relationship was assumed for the out-flux: $J_{out}=K_{\rm off} C_2$, and where the exchange fluxes ($J_{1\rightarrow2}$) are still given by Eqs.\ref{jab},\ref{saturation},\ref{rates}. At steady state, all fluxes must be balanced, including the net flux $J$ between the two compartments: $J_{in}=J_{out}=J$. 

\begin{figure*}
\includegraphics[width=15cm]{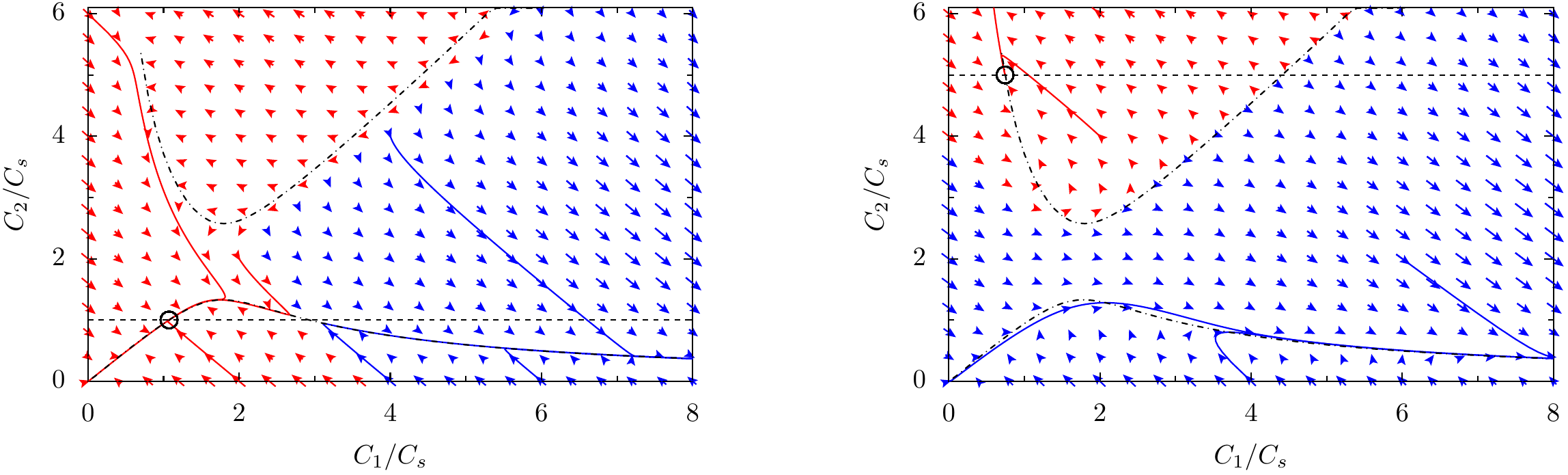}
\caption{\label{phasespace}Phase-space trajectories of system with an exit flux $J_{\rm out} = K_{\rm off} C_2$ ($K_{\rm off}=0.005 K_0$), and an input flux $J_{in} =0.005 K_0 C_s$ (left) and $J_{in} =0.025 K_0 C_s$ (right). Dash-dotted lines represent $\dot{C}_2 = 0$ and dashed lines $\dot{C}_{\rm tot}=0$. Red arrows represent initial condition with convergent trajectories whereas blue arrows are for initial conditions yielding a divergence of $C_1$. }
\end{figure*}

\subsection{Qualitative analysis}\label{sec-open_qualit}

The dynamical behaviour of the set of equations \eq{kineqc} is discussed in some details in Sec.\ref{sec-open_phasespace}, but a qualitative understanding of the open system may be inferred from the results obtained for a closed system. We showed (\fig{closed}b) that the flux $J_{1\rightarrow2}$ cannot exceed a maximum value 
and decreases upon increasing total concentration beyond a threshold. 
In an open system, this behaviour may result in the absence of a steady state: if the influx into compartment $1$ exceeds the maximum net flux from $1\rightarrow 2$, the concentration $C_1$ of the entry compartment steadily increases with time, leading to a further decrease of inter-compartment exchange. In the absence of other compensatory mechanisms, $C_1$ would diverge and $C_2$ would vanish, leading to a vanishing exit flux. This divergence is probably not realistic, but it illustrates the consequence of such non-linear transport for an open system: beyond a critical influx, the system is essentially blocked, filtering transit proteins at a very low flux. While such feature has a negative impact on the rate of transport, it strongly increases the residency time of molecules and may prove advantageous to a system such as the Golgi apparatus, whose function is to process and chemically modify proteins.

\subsection{Phase-space trajectories of an open system}\label{sec-open_phasespace}

We now discuss possible dynamical behaviours of an open systems satisfying the kinetic equation \eq{kineqc}, where the fluxes between the two compartments are given by Eqs.{\ref{jab},\ref{saturation},\ref{rates}. Although the exchange rules between the compartments are symmetric, the existence of external fluxes breaks the symmetry of the system, and different concentrations should be expected in the two compartments even for low incoming flux. The critical behaviour at high incoming flux, as depicted in \fig{closed} for a closed system, has nevertheless a profound impact on the steady states, or the absence thereof.

As discussed above, one expects the flux exchanged between the two compartments to present a maximum value $J_{\rm max}$ (necessarily smaller than the maximum possible flux $K_0 C_s$, see \fig{closed}.b), theoretically leading to a diverging concentration in the first compartment and a vanishing exchange flux if $J_{in}  > J_{\rm max}$. Depending on initial conditions, a non-convergent behaviour might actually also appear for values of $J_{in}$ {\rm a priori} compatible with the existence of a steady-state. For instance, if the initial concentration is very high in the first compartment,  the divergent regime may occur for smaller in-flux $J_{in} < J_{max}$ . This can be understood by considering the phase space trajectories of the system. 

The coordinates in phase space are the concentrations $( C_1 , C_2 )$, and the steady states (if any) are given by the intersections of the $\dot{C}_2=0$ and $\dot{C}_{\rm tot}=0$ curves. Since $J_{\rm out}= K_{\rm off} C_2$, the line $\dot{C}_{\rm tot}=0$ is obviously the line $C_2=J_{in} / K_{\rm off}$, whereas the curve $\dot{C}_2=0$ has to be computed numerically. If these two lines do not intersect, there is no steady state and $C_1$ always diverges. If they do intersect, the thus-defined fixed points may be linearly unstable, or may be surrounded by a basin of attraction, as shown in \fig{phasespace}.

The phase space representation show \fig{phasespace} can be used to study the consequences of a transient change of the input flux (i.e. a pulse or a block of secretion). Let us consider a system which is in a stable steady state $(C_1^1,C_2^1)$ for an input flux $J_{in}^1$. If the incoming flux is changed to $J_{in}^2$ at time $t_1$, the phase space trajectories will be changed, and the system will follow a new trajectory starting from $(C_1^1,C_2^1)$. According to this new trajectory, the system will reach a new position $(C_1^2,C_2^2)$ at a time $t_2$. If the flux is then switched back to its original value $J_{in}^1$,  $(C_1^2,C_2^2)$ will not necessarily be in the attractive region of the stable steady state. Therefore, a transient change of the incoming flux may bring the system out of a stable steady state.
In the case of a strong pulse ($J_{in}^2 >> J_{in}^1$) the system may follow a divergent trajectory and the concentration $C_1$ will increase strongly with time. Formally, whatever the (finite) value of $(C_1^2, C_2^2)$ after a pulse, the system may reach a stationary regime if the incoming flux $J_{in}$ after the pulse is small enough. However, this may take a very long time. The approximation $C_1 \rightarrow \infty$ , $J_{in} = 0$ shows this time grows like ${(C_1^2)}^2$.

\section{Consequence of cooperative transport for protein maturation and sorting}\label{sec-sorting}

We now quantify the consequences of the kind of cooperative transport considered here on protein maturation and sorting. We investigate the situation 
sketched in \fig{open}, where a molecular species $A$ enters the system via compartment $1$ and is transformed into a species $B$ by maturation enzymes, before leaving the system via compartment $2$. The processing accuracy is defined as the total fraction of the input that exits the system as mature ($B$) molecules: 
\be
{\rm Accuracy}\equiv\frac{1}{A_0}\int _0^\infty J_{out}^Bdt
\label{accuracy}
\ee
where $A_0=\int_0^\infty J_{in}dt$ is the total amount of $A$ molecules to have entered the system and $J_{out}^B$ is the out-flux of $B$ molecules. The accuracy thus defined  reaches unity when no molecules exit the system without being processed ($J_{out}^A=0$).

A Michaelis-Menten maturation kinetics is chosen in order to account for the limited amount of enzymes in the system. Calling $A_1$ and $B_1$ the concentrations of $A$ and $B$ in the first compartment, we have:
\be
\partial_t B_{1}=R(A_{1}) A_{1}=R_0 C_m\frac{A_{1}}{A_{1}+C_m}
\label{dotb}
\ee
with an identical kinetics in compartment 2. Here,
$R_0$ is the maximal maturation rate and $C_m$ is the concentration of $A$ beyond which enzymatic reaction saturates. For simplicity, we assume that the state ($A$ or $B$) of a molecule influences neither its transport between compartments nor its export from the system,  so that \eq{kineqc} is still valid for the concentrations $C_{1,2}=A_{1,2}+B_{1,2}$. Taking the weights of $A$ and $B$ in the fluxes to be their respective weights in the compartments :
\begin{subequations}
\begin{align}
J^A=\frac{A_1}{A_1+B_1}J_{1 \rightarrow 2}-\frac{A_2}{A_2+B_2}J_{2 \rightarrow 1}
\\ J^B=\frac{B_1}{A_1+B_1}J_{1 \rightarrow 2}-\frac{B_2}{A_2+B_2}J_{2 \rightarrow 1},
\end{align}
\label{bigkineq1}
\end{subequations}
the following set of kinetic equations is obtained:
\begin{subequations}
\begin{align}
\dot A_1=J_{in}-R(A_1) A_1-J^A
\\ \dot B_1=R(A_1) A_1-J^B
\\ \dot A_2=-R(A_2) A_2+J^A-K_{\rm off} A_2 
\\ \dot B_2=R(A_2) A_2+J^B-K_{\rm off} B_2
\end{align}
\label{bigkineq2}
\end{subequations}

\begin{figure}[t]
\includegraphics[width=7cm]{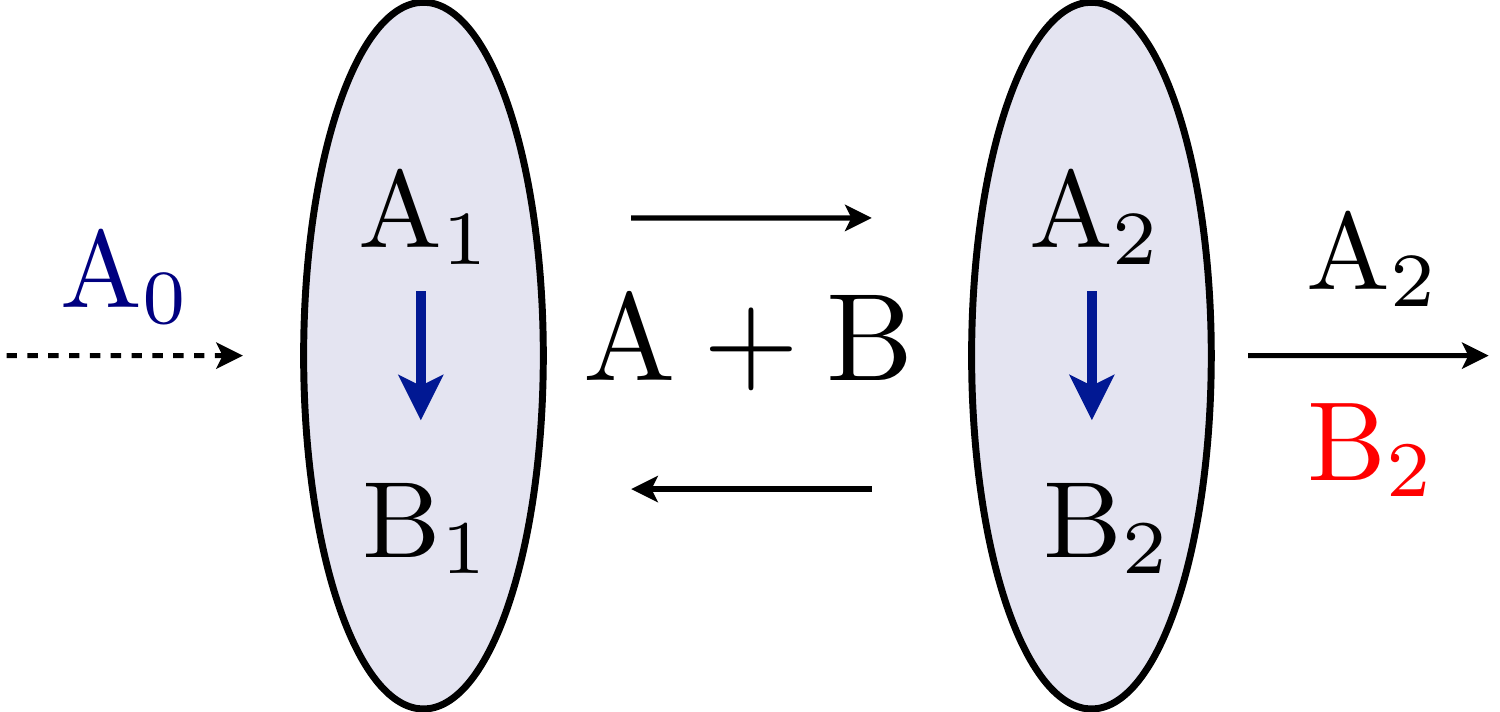}
\caption{\label{open} Sketch of an open system with protein maturation. Particle enter the system through compartment $1$, undergo maturation $A\rightarrow B$ while in the system, are exchange between compartment via cooperative transport, and exit the system via compartment $2$.}
\end{figure}

Normalizing rates with the vesiculation rate $K_0$ and concentrations with the concentration $C_s$ at which secretion saturates, \eq{bigkineq2} is controlled by $5$ parameters. These are: $r_0=R_0/K_0$ and $C_m/C_s$, which compare the activity of the maturation enzymes and of the secretion machinery, $C_f/C_s$, which defines the threshold for dominant specific fusion (\eq{cstar}), and $K_{\rm off}/K_0$, which compares exit and exchange rates. The fifth parameter is the normalized amount of material going through the system: $A_0/C_s$. For simplicity, we investigate a situation similar to the so-called {\em pulse-chase} procedure\cite{trucco:2004}, where a fixed amount of material  is delivered to the system in a finite amount of time (which we assume very small), and set $A_1(t=0)=A_0$ and $J_{in}=0$ below.

In order to focus on the role of cooperative transport, we further assume that particle export is not a rate-limiting step ($K_{\rm off}/K_0\rightarrow\infty$), and we analyze  the processing accuracy in terms of a competition between the kinetics of maturation and transport (controlled by 4 parameters). 

\subsection{Processing accuracy for linear transport}\label{sec-sorting_linear}

In order to quantify the consequences of cooperativity on the processing accuracy of a two-compartments system, we compute the accuracy of a perfectly linear system by linearizing Eqs.{\ref{jab},\ref{saturation},\ref{rates} when $A_0\ll C_m,C_s,C_f$, yielding: $J_{1 \rightarrow 2} = K_0 C_1/2$ and $J_{2 \rightarrow 1} = K_0 C_2/2$. Choosing a linear exit flux for simplicity ($J_{\rm out}=K_{\rm off} C_2$, for which we later take the limit $K_{\rm off} \rightarrow\infty$), and the initial conditions $C_1(t=0)=C_1(0)$ and $C_2(t=0)=0$, the kinetic evolution of the vector ${\boldsymbol{C}}=\{C_1(t),C_2(t)\}$ is easily obtained:
\begin{eqnarray}
\boldsymbol{C} (t) = e^{M_l t} \left( \begin{array}{c} C_{1}(0) \\ 0 \end{array} \right)\hspace{0.1cm},\ 
 M_l= -\frac{K_0}{2} \left[ \begin{array}{cc} 1 &  1 \\ 1 &  1 +2 k_{\rm off}  \end{array} \right]  \label{eqlin}\ \ \ \ \ 
\end{eqnarray}
where $k_{\rm off} = \frac{K_{\rm off}}{K_0}$. 
The matrix $M_l$ can be diagonalized, and the matrix exponential becomes a regular exponential, and the concentration in the second compartment reads:
\begin{equation}
C_2 (t) =\frac{ C_1(0)}{2 \sqrt{1+k_{\rm off}^2}} \left(e^{\alpha_+t} -e^{\alpha_- t}\right)
\end{equation}
with the  eigenvalues : 
\begin{eqnarray}
\alpha_\pm=\frac{K_0}{2} \left( \pm \sqrt{1+k_{\rm off}^2}-(1+k_{\rm off})\right)
\label{eigenvalues}
\end{eqnarray}

The (normalized) probability density that a particle exits the system from the second compartment at time $t$ is $P_{exit}(t) = K_{\rm off} C_2(t)/C_1(0)$: 
\begin{eqnarray}
P_{exit}(t)=\frac{ K_ 0k_{\rm off}}{\sqrt{1+  k_{\rm off}^2}} \left(e^{\alpha_+t} -e^{\alpha_- t}\right)\label{pexitdet}
\end{eqnarray}
The mean residence time of a particle in the system is thus  $\langle T\rangle\equiv\int_0^\infty dt (t P_{exit}(t))=2(1/K_0+1/K_{\rm off})$.

The accuracy of protein maturation ($A \rightarrow B$) and sorting is defined as the fraction of the total quantity of molecules that entered the system to leave as matured molecules (\eq{accuracy}). It may also be written as: 
\begin{equation}
{\rm Accuracy} \, = \int_0^{+\infty}  P_{exit}(t) P(B,t|A,0) dt \label{defeffic}
\end{equation}	
where which $P(B,t|A,0)$ is the probability for a molecule to be mature (state $B$) at time $t$ while starting immature (state $A$) at $t=0$. At the linear level, the maturation kinetics (\eq{dotb}) becomes:  $\partial_t B_\alpha=R_0 A_\alpha$, and $P(B,t|A,0)=1-e^{-R_0 t}$. The efficiency of the linear system may then be computed analytically using Eqs.(\ref{eigenvalues},\ref{pexitdet},\ref{defeffic}), yielding :
\be
{\rm Accuracy}|_{\rm linear}\, =\frac{ 2 r_0\left(1+ r_0+ k_{\rm off}\right)}{ k_{\rm off}+ 2 r_0 \left(1+ r_0+ k_{\rm off}\right)}
\label{efflin}
\ee
with $r_0=R_0/K_0$. Taking the limit $k_{\rm off}\rightarrow\infty$ as discussed, the benchmark to which more complex transport and maturation processes must be compared is thus the linear accuracy: ${\rm Accuracy}|_{\rm linear}\rightarrow2r_0/(1+2r_0)$. 

\subsection{Processing accuracy without specific vesicular fusion}\label{sec-sorting_randomfusion}

Saturation of maturation enzymes and transport intermediates ($C_m,C_s<A_0\ll C_f$, with $A_0$ the initial particle concentration) has mixed effects on the systems processing accuracy. Saturation of inter-compartment transport at high concentration (for $A_0 \gg C_s$) causes the particle residency time of molecules  to grow as $A_0$ (\eq{saturation}) while saturation of enzymatic reaction (for ($A_0 \gg C_m$) causes the mean maturation time increase linearly with $A_0$ (\eq{dotb}), so the net effect on processing accuracy depends on the precise values of the parameters. 

In order to get a fell for the role of the different parameters, we compute the first order correction  to the linear processing kinetics studied in Sec.\ref{sec-sorting_linear}, in the limit  of very fast exit from the second compartment: $k_{\rm off} \rightarrow \infty$. In this case,  the accuracy is  controlled by the flux exiting the first compartment, now written $J_{\rm out} = \frac{1}{2} K_0 C_s S(C_1)$ and \eq{bigkineq2} may be rewritten:
\begin{subequations}
\begin{align}
C = A + B \\
\dot{A} = - C_s \frac{1}{2} \frac{A}{C+C_s} - r_0 C_m \frac{A}{A+C_m} \\
\dot{B} = - C_s \frac{1}{2} \frac{B}{C+C_s} + r_0 C_m \frac{A}{A+C_m}
\end{align}
\label{kineqlin2}
\end{subequations}
where the subscript $1$ has been dropped in the concentrations, time has been normalized by  $1/K_0$, and $r_0\equiv R_0/K_0$. Taylor expansion of this set of equation for $A_0 \ll min(C_s,C_m)$ yields the first order correction to the accuracy of the linear system (\eq{efflin}):
\begin{eqnarray}
\begin{split}
{\rm Accuracy} = \frac{2 r_0}{1+2r_0}& +\frac{A_0}{C_s} \frac{C_m (1+2 r_0) - C_s (1 + r_0)}{C_m (1+r_0) (1+2r_0)^2}
\\  &+{\cal O}\left[ \left(\frac{A_0}{C_s} \right)^2 \right]
\end {split}
\end{eqnarray}

An increase in processing accuracy at high concentration requires that maturation saturates after secretion according to  $C_m/C_s>(1+r_0)/(1+2 r_0)$. This can be seen in \fig{fig-accuracy}, which shows the variation of the system's processing accuracy as a function of the total amount of material to be processed, in the absence of cooperative fusion (dashed lines).

\begin{figure}[t]
\includegraphics[width=\columnwidth]{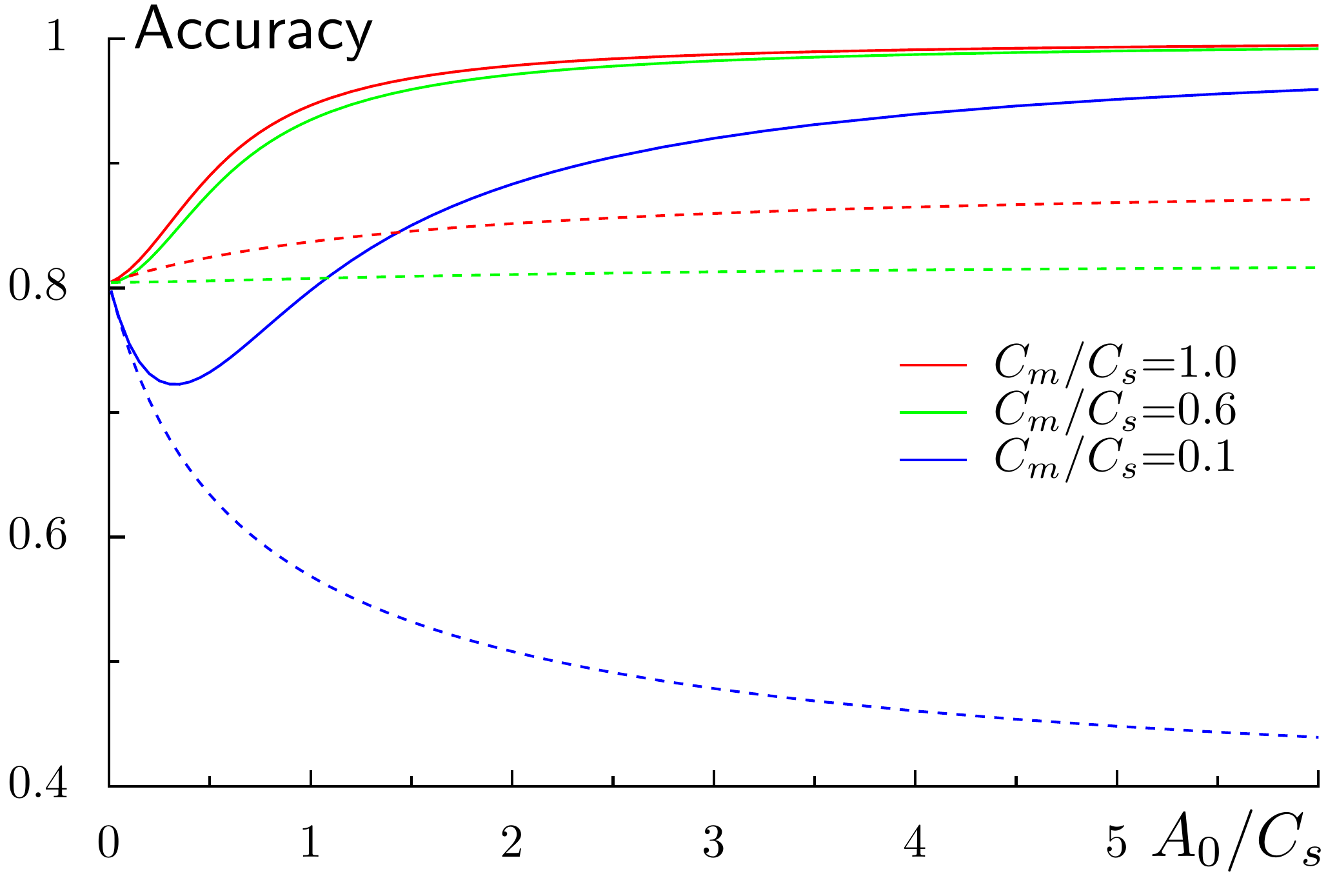}
\caption{\label{fig-accuracy} Accuracy (\eq{accuracy}) as a function of the initial concentration $A_0$, for different saturation ratios $C_m/C_s$ of maturation and transport. At high concentration, specific vesicle fusion greatly enhance processing accuracy (solid lines, with $C_f/C_s=0.1$), as compared to random fusion ($C_f\rightarrow\infty$, dashed lines) (with $R_0/K_0=2$, and $K_{\rm off}/K_0=100$).}
\end{figure}

\subsection{Processing accuracy and cooperative transport}\label{sec-sorting_full}

Cooperative fusion, when combined with saturation of the transport, leads to a robust increase of the processing accuracy of a compartmentalized organelle responsible for protein maturation and sorting  (see \fig{fig-accuracy}, solid lines). This increase can be understood as follows; at high concentration ($A_0>C_f,C_s$), specific interactions promote backward fusion of vesicles secreted by the highly concentrated compartment. As the forward fusion probability is very low ($P_{1\rightarrow2}\sim 1/A_0$, \eq{rates}) the mean residency time increases as $A_0^2$, as compared to the linear increased observed in the absence of specific fusion (Sec.\ref{sec-sorting_randomfusion}). On the other hand,  the mean maturation time is still linear in $A_0$, so high concentrations lead to a more pronounced increase of  the residency time compared to the maturation time, resulting in an increased processing accuracy at high concentration, even if the chemical transformation is performed by a limited amount of maturation enzymes ($C_m\ll C_s$).

\section{Conclusion and outlook}

The predicted high processing accuracy displayed \fig{fig-accuracy} essentially stems from the increase of the residency time of molecules transiting through the system.  In striking contrast with the usual Fick's law of gradient-driven transport,  
cooperative transport through the compartmentalized system described here is strongly impaired by large concentration gradients. A strong prediction of our model is that the transport time actually increases with an increasing incoming flux (above a threshold). 
Pulse-chase experiments on the Golgi seem to show this trend, but data are still too scarce for a direct comparison (see Fig.4.$l$ in \cite{trucco:2004}). Although an apparent functional  drawback, slow transport through organelles is common. For instance, the typical transport time across the Golgi is  of order of $20$ minutes \cite{patterson:2008}, whereas diffusion of a membrane protein over an area equal to that of the entire Golgi apparatus (of order $10\mu$m$^2$) should be of order one minute (with a diffusion coefficient $\sim 0.1\mu$m$^2/$s \cite{lippincott:2000}).

In this paper, we showed that organelles constantly exchanging material via transport vesicles may spontaneously adopt different biochemical identities, provided: {\em i)} the flux of vesicles secreted by an organelle is bounded, and {\em ii)} there exists  some level of specific vesicle-organelle fusion  directed by molecular recognition. In open systems traversed by fluxes, these transport properties give rise to a dynamical switch from a linear to a low throughput kinetics above a critical influx. For compartmentalized organelles whose function is to process and export influxes of proteins, such as the Golgi apparatus, this switch allows the export rate to spontaneously adjust to the amount of material to be processed, a definitive functional advantage that may avoid the release of unprocessed material even under high influx.  Future extension of the present model to multi-component transport  (see S.I.) will allow to assess the importance of the exchange of membrane area (lipids) between compartments (the compartment sizes were assumed constant here). 
It will also allow us to explore the role of specialized  membrane domains; domains for protein processing and domains for protein export, which have recently been reported \cite{patterson:2008}.

\begin{acknowledgments} We gratefully acknowledge M. Rao, S. Mayor and N. Gov for stimulating discussion and are indebted to  R. Phillips for critical reading of the manuscript.  \end{acknowledgments}

\bibliographystyle{apsrev4-1}

%

\appendix

\begin{figure*} 
\section{Numerical simulation - Finite vesicle fusion time}

To test the predictions of Eqs.\ref{finitevestime1},\ref{finitevestime2}, in the case vesicle fusion occurs with a finite rate, we performed a numerical simulation of a system with finite vesicle fusion time and a total concentration $C_{\rm tot}$ and compared the location of the critical line with the infinitely fast fusion model, the equations of which we solved numerically. The numerical simulation consists of two compartments of concentrations $C_1$ and $C_2$ from which vesicles may bud with a rate $K_v$. Each vesicle budding from a compartment $\alpha$ has a saturation $S(C_{\alpha})$. At each timestep, each vesicle may merge with a compartment at a rate $W_r$, and the compartment is chosen according to the probability $P$ described in the main text. The algorithm may be written as follows :

\begin{small}
\lstset{language=Python}
\begin{lstlisting}
#Fusion probability of a vesicle of saturation Sv with the first compartment
def Pf1(Sv,C1,C2)=(Sv*C1+Cf)/( 2*Cf + Sv*(C1+C2) )
#Saturation of the vesicles leaving from a compartment of concentration C
def S(C) = C / (C+Cs)

#Sves[i]: Saturation of the i-th vesicle
#Nves : number of vesicles

while t<Tmax :
	t=t+dt
	
	#Checks if a vesicle leaves the first compartment
	if rand(1) < Kv*dt :
		Nves=Nves+1
		Sves[Nves]=S(C1)
		C1=C1 - Cv*Sves[Nves]

	#Checks if a vesicle leaves the second compartment	
	if rand(1) < Kv*dt :
		Nves=Nves+1
		Sves[Nves]=S(C2)
		C2=C2 - Cv*Sves[Nves]
		
	#Checks for each vesicle if it merges with a compartment
	for i=1 to Nves :
		if rand(1) < Wr*dt :
			if rand(1) < Pf1(Sves[i],C1,C2) :
				C1=C1+Cv*Sves[i]
			else :
				C2=C2+Cv*Sves[i]
			Sves[i]=0
	reorder(Sves,Nves)		
\end{lstlisting}

\end{small}
\end{figure*}


\end{document}